\documentclass[12pt,a4paper]{article}
 
\usepackage[scale={.75,.8}]{geometry}
\usepackage{pstricks}
\usepackage{graphicx}
\usepackage[latin2]{inputenc}
\usepackage{epic}
\usepackage{latexsym}
\usepackage{epsf}
\usepackage{epsfig}
\usepackage{cite}
\usepackage{amssymb,amsmath}

\newcommand{\eref}[1]{eq.~(\ref{e.#1})} 
\newcommand{\erefn}[1]{(\ref{e.#1})}
 
\newcommand{\aref}[1]{\ref{a.#1}}
\newcommand{\sref}[1]{Section~\ref{s.#1}}
\newcommand{\cref}[1]{Chapter~\ref{c.#1}}

\newcommand{\nl}{& \nonumber \\ &}

\def\ds{\displaystyle}

\def\kahler{K\"ahler\hspace{0.1cm}}
\def\sugra{supergravity\hspace{0.1cm}}

\def\beq{\begin{equation}} 
\def\eeq{\end{equation}} 
\def\bea{\begin{eqnarray}}  
\def\eea{\end{eqnarray}}  
\def\ba{\begin{array}}  
\def\ea{\end{array}}   
\def\bi{\begin{itemize}}  
\def\ei{\end{itemize}}  
\def\be{\begin{enumerate}}  
\def\ee{\end{enumerate}}  
\def\beq{\begin{equation}}  
\def\eeq{\end{equation}}  
\def\bc{\begin{center}}
\def\ec{\end{center}}

\def\cl{{\mathcal L}}

\def\co{{\mathcal O}}

\def\ra{\rangle} 
\def\la{\langle}  

\def\pa{\partial}  
\def\re{{\rm Re} \,}
\def\im{{\rm Im}\,}

\def\rt{\sqrt{2}}

\def\bn{{(n)}}

\def\hc{{\rm h.c.}}

\def\ZZ{\mathbb{Z}}    
\def\ov{\overline}

\def\lb{\label}
\def\eps{\epsilon}

\begin{document}

\pagestyle{empty}
\begin{flushright}
hep-th/0504091\\
DESY-05-056
\end{flushright}
\vskip 1.5cm

\begin{center}
{\huge Gravity mediated supersymmetry breaking in six dimensions}
\end{center}
\vspace*{5mm} \noindent
\vskip 0.5cm
\centerline{\bf Adam Falkowski ${{}^{a,b}}$, Hyun Min Lee${{}^{b}}$, Christoph L\"udeling ${{}^{b}}$}
\vskip 1cm
\centerline{\em  ${{}^{a}}$ Institute of Theoretical Physics, Warsaw University}
\centerline{\em ul.\ Ho\.za 69, PL-00-681 Warsaw, Poland}
\vskip 0.5cm
\centerline{\em  ${{}^{b}}$ Deutsches Elektronen-Synchrotron DESY}
\centerline{\em Notkestra\ss e 85, 22607 Hamburg, Germany}
\vskip .5cm
\centerline{\tt \small 
adam.falkowski@desy.de, hyun.min.lee@desy.de, christoph.luedeling@desy.de}
\vskip1cm
\centerline{\bf Abstract}
\vskip .3cm

We study gravity mediated supersymmetry breaking in four-dimensional effective theories derived from six-dimensional brane-world supergravity.
Using the Noether method we construct a locally supersymmetric action for a bulk-brane system consisting of the minimal six-dimensional supergravity coupled to  vector and chiral multiplets located at four-dimensional branes.
Couplings of the bulk moduli to the brane are uniquely fixed, in particular, they are flavour universal.   
We compactify this system on $T_2/\mathbb{Z}_2$ and derive the four-dimensional effective supergravity.
The tree-level effective \kahler potential is not of the sequestered form, therefore gravity mediation may occur at tree-level.
We identify one scenario of moduli stabilization in which the soft scalar masses squared are postive.

\vskip .3cm

\setcounter{page}{0}
\newpage
\pagestyle{plain}

\section{Introduction}

Supersymmetry breaking and its mediation to the observable sector is an important issue for any supersymmetric extension of the Standard Model. In the Minimal Supersymmetric Standard Model 
supersymmetry breaking is parametrized by a set of soft terms: gaugino, squark, slepton and Higgs  masses and trilinear A-terms. But, the allowed pattern of soft terms is strongly constrained by the observed features of the low-energy physics. Therefore an underlying theory of soft terms generation is needed. 

Gravity mediation \cite{gm} is a simple and economical possibility  because it is a generic consequence of local supersymmetry. 
Let us denote the three generations of visible matter superfields by $Q^i$ and assume supersymmetry breaking is parametrized by an $F$-component of a chiral superfield $\Sigma$.  The four-dimensional (4d) supergravity action is specified by three functions: the \kahler potential $K= - 3 \log \Omega$, the  superpotential $W$ and the gauge kinetic function $f$. 
In this paper we concentrate on soft scalar masses and these are encoded in $\Omega$.   
Quite generally, the \kahler potential can be expanded in powers of $|Q|^2$, 
\beq
\Omega = \Omega_0(\Sigma^\dagger,\Sigma) 
-  {1 \over 3 M_p^2}  Q^i{}^\dagger Q^j \left (
\delta_{i j} + C_{i j}(\Sigma,\Sigma^\dagger) \right )+ \co (Q^4) \,.
\eeq 
Without loss of generality we can assume $C_{i j}(\la \Sigma \ra,\la \Sigma^\dagger \ra) = 0$.  
The functions $C_{i j}$ are called the contact terms as they control supersymmetry breaking mediation to the visible scalar sector. Indeed, in a vacuum with the vanishing cosmological constant the supergravity $F$-term potential contributes to soft scalar masses as  
\beq
(m_Q^2)_{i j} = - |F_\Sigma|^2 
\left ( \pa_\Sigma \pa_{\ov \Sigma} C_{i j} 
- \pa_\Sigma C_{i k} \pa_{\ov \Sigma} C_{k j} \right )
\, .
\eeq     
But in order to avoid excessive flavour changing neutral currents we need $(m_Q^2)_{i j}$ to be approximately diagonal (in the basis in which quark and lepton masses are diagonal). To this end we need the contact terms of a very special form,  
either universal, $C_{i j} \approx  \delta_{i j}$,  or  aligned with  the Yukawa matrices. 
 To justify this ad-hoc assumption it is desirable to have a `theory  of contact terms'. That is, 4d supergravity should be embedded in a more fundamental set-up in which the contact terms could be actually computed and the universalness (or alignment) could be explained.   

Higher dimensional brane-world supergravities offer promising candidates for such a theory of contact terms  \cite{rasu0}.  
First of all, the supersymmetry breaking sector can be spatially separated from the visible sector. 
This is achieved by assuming that visible matter fields are confined to a 4d brane (the visible brane) while supersymmetry breaking dynamics  takes place on a different brane (the hidden brane).
The couplings between the two sectors that could arise from integrating out heavy fields from the UV completion of the theory can be neglected if separation between the branes is large enough (in units of the fundamental scale). 
In such case, the couplings arise only from integrating out Kaluza-Klein (KK) modes of bulk degrees of freedom and are calculable within the field theoretical framework. 
Still there is no guarantee that the resulting soft terms are flavour blind, as the bulk may contain matter fields coupling non-universally to SM generations.
However, the degrees of freedom belonging to the higher dimensional gravity multiplet necessarily couple universally.
If the corresponding moduli dominate supersymmetry breaking mediation (that is, either bulk matter fields  are absent or their F-terms are negligible) we end up with calculable and flavour blind soft terms.  
Of course, the flavour problem can be really solved only if the issue of generating Yukawa hierarchy in the visible sector is also addressed. 
If the correct Yukawa pattern  is generated at the fundamental scale then the non-universal corrections to soft terms can be neglected, provided the compactification scale is well below the fundamental scale.  
But when flavour structures are generated at low energies by the Froggatt-Nielsen mechanism one indeed expects non-negligible contributions to A-terms that could induce dangerous FCNC \cite{rovi}.

Concrete realizations of this idea were extensively studied in the settings of 5d supergravity compactified on $S_1/\mathbb{Z}_2$ \cite{lusu,bugago,ghri,rascst}. 
It turns out that the issue of supersymmetry breaking mediation can be most conveniently studied within the 4d  effective theory after all Kaluza-Klein modes are integrated out. 
In the minimal version with only gravity in the bulk such effective theory  contains only one modulus $T$ corresponding to the size of the fifth dimension. Apart from that there are visible and hidden brane matter fields, $Q^i_V$ and $Q_H$.  In the simplest scenario, when the bulk cosmological constant vanishes, the \kahler potential of the effective 4d supergravity at tree-level is of the no-scale form:  
\beq 
\lb{e.o5t0}
\Omega_{5d} = {1 \over 2} (T + \ov T) 
- {1 \over 3}\Omega_V(Q^i_V,Q^i_V{}^\dagger)  - {1 \over 3} \Omega_H (Q_H, Q_H^\dagger) \, .
\eeq
In this \kahler potential sequestering is explicit - the contact terms between visible and hidden matter  are absent. 
The $T$ modulus does not couple to the visible matter as well, therefore in the 5d scenario there is no tree-level gravity mediation.   
The contact terms are however generated at loop level after  one loop corrections from  Kaluza-Klein modes of the gravity multiplet are included. Then supersymmetry breaking can be  mediated to the visible brane but, 
unfortunately, the sign of these gravity mediated soft masses squared is negative. Therefore the minimal 5d set-up does not provide for a phenomenologically viable theory of contact terms. Generalizing the set-up to warped supergravity does not improve the situation \cite{grrasc,aa}. 
{ Positive soft scalar masses can be however obtained in models with sizable brane gravity kinetic terms \cite{rascst,grrasc}.}  
  
In this paper we extend the previous studies to 6d supergravities compactified on $T_2/\mathbb{Z}_2$. Using the Noether method \cite{noether} we construct a locally supersymmetric action for a bulk-brane system containing the minimal $N=2$ 6d supergravity and $N=1$ chiral and vector multiplets confined to a 4d brane at an orbifold fixed point. 
{It should be stressed that our approach is purely field-theoretical. Embedding this set-up in string theory would presumably result in additional constraints on the spectrum and couplings, but this issue is beyond the scope of this paper. Our results hold irrespectively of which string theory provides a UV completion to our model, as long as local supersymmetry and 6d locality remain valid below the string scale.}   

 After deriving bulk-brane couplings  we identify the 4d effective supergravity describing the dynamics of this system at energies below the compactification scale. Except for the visible and hidden brane matter fields, $Q_V^i$ and $Q_H$, the effective theory contains three moduli $T$, $S$ and $\tau$. The \kahler potential is not of the sequestered form:     
\beq 
\label{e.introomega}
\Omega_{6d} = {1 \over 2}
\left (T + \ov T 
- 2\Omega_V(Q^i_V,Q^i_V{}^\dagger)  - 2  \Omega_H(Q_H,Q_H^\dagger) \right)^{1/3}
\left (S + \ov S\right)^{1/3}  \left (\tau + \ov \tau\right)^{1/3}
\eeq
Expanding this \kahler potential in powers of $|Q_V|^2$ we find the contact terms involve not only the moduli but also the hidden brane fields. Therefore this \kahler potential allows for both  moduli and brane-to-brane mediation of supersymmetry breaking at tree-level. 

Absence of sequestering in higher dimensional brane world models was already discussed by Anisimov et al. in ref. \cite{andigr}. 
These authors argue that higher dimensional locality does not have to  show up below the compactification scale.  
The reason is that brane world models typically involve bulk fields with non-trivial coupling to the branes. 
Decoupling  KK modes of these bulk fields may lead to seemingly `non-local' contact interactions involving  fields from different branes.
These operators are suppressed by the volume of the extra dimensions, but the gravitino mass, which sets the scale of supersymmetry breaking, is suppressed by the same factor. 
The conclusion of ref. \cite{andigr} is that in generic brane world models contact terms arise at tree-level leading to (in general non-diagonal) soft masses of order the gravitino mass.

Non-sequestering is indeed a feature of our \kahler potential \erefn{introomega}. 
We will argue however that  the situation is slightly more subtle. 
Compactification of the 6d action we consider does not contain `non-local' operators (except for vector-vector interactions that are also present in the sequestered case and play no role in supersymmetry breaking). 
More precisely, scalar-scalar `non-local' interactions are present in the off-shell formulation, but vanish when the auxiliary fields are integrated out. 
This conclusion relies not only on the form of the \kahler potential \erefn{introomega}, but also on the fact that the superpotential describing the compactified action does not depend on the $T$ modulus.    
As a consequence,  tree-level supersymmetry breaking mediation does not operate, in spite of the non-sequestered form of \eref{introomega}. 
`Non-local' interactions and tree-level mediation  does however occur when the minimal 6d action is supplemented by certain higher order operators.  
In the effective 4d description these higher order operators induce dependence of the superpotential on the $T$ modulus. 
Thus our conclusion is  that information about sequestering in brane world models is encoded not only in the contact terms of the \kahler potential, but also in  the structure of  the superpotential.

This paper is organized as follows. In \sref{bba} we construct minimal 6d supergravity coupled to matter on codimension two branes. In \sref{lees} we derive the \kahler potential, superpotential and  gauge kinetic function of the resulting low energy 4d supergravity. In \sref{gmsb} and \sref{ms} we discuss supersymmetry breaking mediation and moduli stabilization and in \sref{c} we present our conclusions.

\section{Bulk-brane action}
\label{s.bba}

In this section we construct a locally supersymmetric action for the 6d minimal $N=2$ (eight supercharges)  supergravity coupled to $N=1$ chiral and vector multiplets that are confined to a 4d brane at an orbifold fixed point. The construction is done for just one chiral and one abelian vector multiplet with minimal kinetic terms, but it can be easily extended to more general cases. 
We use the Noether method, which is a simple but very efficient way to derive the bulk-brane couplings. Starting with a locally supersymmetric bulk action and  a globally supersymmetric brane action we systematically add new terms to the action and supersymmetry transformations until the whole set-up becomes locally supersymmetric.
 We  work out all necessary zero- and two-fermion terms so that all  two-fermion supersymmetric variations  of the bulk-brane action cancel. It turns out that  the analysis at the two-fermion level is sufficient to read off  unambiguously the form of the 4d low energy effective supergravity, which is our main objective in this paper. Therefore we stop our procedure at this point and do not  derive four-fermion terms in the brane action. This means that we tacitly assume that our construction can be completed such that all supersymmetric variations cancel and supersymmetry algebra closes on-shell.

The starting point is to  write down the bulk action. 
The minimal set-up contains  6d  $N=2$ gravity+tensor multiplet that includes a sechsbein
$e_A^a$, a gravitino $\psi_A$, a Kalb-Ramond  form $B_{A B}$ with a three-form field strength
$H_{A B C} = 3 \pa_{[A} B_{B C]}$, a fermionic dilatino $\chi$ and a real scalar $\Phi$ called
the dilaton. 
The indices  $A,B,C,a$ run over  $0 \dots 3,5,6$ and the fermions obey 6d chirality conditions. 
The conventions used in this paper are collected in \aref{nc6}.
The bulk action up to four-fermion terms reads \cite{masc,nise}   
\begin{align} \label{e.6da}
  \begin{split}
    \cl_{\rm bulk} &= M_6^4 e_6 \left[\frac{1}{2} R_6 - i \ov{\psi_A} \Gamma^{A B C} D_B \psi_C +
      \frac{1}{12}e^{- 2 \Phi} H_{A B C} H^{A B C} + i \ov{\chi} \Gamma^{A} D_A \chi \right.\\
    & \mspace{75mu} + \frac{1}{2} \pa_A \Phi \pa^A \Phi 
      - \frac{i}{12 \sqrt{2}} \ov{\psi_A} \Gamma^{A B C D E} \psi_B e^{- \Phi} H_{C D E} \\
    & \mspace{75mu}- {i \over 2 \sqrt{2}} \ov{\psi_A} \Gamma^{C} \psi_B  e^{- \Phi}  H_{A B C} 
      +{1 \over 12 \sqrt{2}}\ov{\psi_A} \Gamma^{A B C D}\chi e^{- \Phi} H_{B C D}+ \hc \\
    & \mspace{75mu} -{1 \over 4 \sqrt{2}}\ov{\psi_A} \Gamma^{B C}\chi e^{- \Phi} H_{ABC} +\hc\\
    & \mspace{75mu}\left.+ {i \over 12 \sqrt{2}}  \ov \chi  \Gamma^{A B C} \chi  e^{- \Phi}
      H_{A B C} - {1 \over 2} \ov \chi \Gamma^{A}  \Gamma^{B} \psi_A \pa_B \Phi + \hc \right]
    \, ,
  \end{split}
\end{align}
while the supersymmetry transformations, up to three-fermion terms are given by
\begin{align}
\label{e.6ds}
  \delta e_A^a &= {i \over 2}  \ov{\psi_A} \Gamma^a \epsilon + \hc \, , \notag\\
  \delta \psi_A &= D_A \eps + {1 \over 24 \sqrt{2}}(\Gamma_{A B C D} - 3 g_{A B} \Gamma_{C D})
  \eps \, e^{- \Phi} H_{B C D}  \, , \notag\\
  \delta B_{A B}  &= {i \over \sqrt 2} e^{\Phi} \ov\psi_{[A} \Gamma_{B]} \epsilon - {1 \over 2
    \sqrt 2} e^{\Phi} \ov \chi \Gamma_{A B} \epsilon +  \hc \, , \notag\\
  \delta \chi &= - {i \over 2} \Gamma^A \epsilon \pa_A \Phi + {i \over 12 \sqrt{2}} \Gamma_{A B
    C}\eps e^{- \Phi} H_{A B C} \, , \notag\\
  \delta \Phi &=  {1 \over 2} \ov \epsilon \,  \chi + \hc \, .
\end{align}

{The model as it stands is inconsistent because of gravitational anomalies. 
Extension to anomaly-free spectrum is possible \cite{rasase}, but it requires a large number of additional bulk multiplets 
(gravitational anomalies vanish when  $244 + n_V - n_H - n_T = 0$, where $n_{V,H,T}$ denote the number of vector, hyper- and tensor multiplets, respectively \cite{er}). }
{ It is of course not possible to give 6d masses to all of these multiplets and their zero modes will survive below the compactification scale.  
It is however conceivable that these multiplets turn out to be irrelevant for the issue of supersymmetry breaking mediation. 
This is the case when they are stabilized by low energy 4d dynamics such that their masses are much larger than the supersymmetry breaking scale and all F-terms in this sector are negligible. 
We will make this ad hoc assumption in what follows and we continue working with the minimal set-up.}

Next, we need to define the orbifold action of $\mathbb{Z}_2$ in a way consistent with the 6d action. 
The torus $T_2$  is parametrized by  $x_5,x_6 \in (-\pi R, \pi R]$
 and the   $\mathbb{Z}_2$ acts as  \linebreak[4] \mbox{$(x_5,x_6) \to (-x_5,-x_6)$}. 
There are four orbifold fixed points: $(0,0)$, $(0,\pi R)$, $(\pi R,0)$ and  $(\pi R,\pi R)$.
We choose the field components
 $e_\mu^m$, $e_\alpha^\beta$, $B_{\mu\nu}$, $B_{\alpha\beta}$, $\phi$ 
 $\psi_\mu^+$, $\psi_\alpha^+$ and $\chi^+$ to be even under $\mathbb{Z}_2$, 
$f(-x_5,-x_6) = f(x_5,x_6)$, 
while $e_\mu^\alpha$, $e_\alpha^m$, $B_{\mu\alpha}$, 
$\psi_\mu^-$, $\psi_\alpha^-$ and $\chi^-$ to be odd,  
$f(-x_5,-x_6) = -f(x_5,x_6)$. 
Here $\mu,\nu,m = 0 \dots 3$. See \aref{nc6} for definitions of the fermionic components.

We move to the Noether construction of the brane action. In fact, the procedure we employ here is very similar to that of coupling 4d brane to 5d supergravity \cite{aa} or 10d brane to 11d supergravity \cite{howi}. It can be summarized in the following steps:
\be 
\item {\it Global brane action.}  Here we write a globally supersymmetric action for  
 a chiral multiplet $(Q,\psi_Q)$   charged under 
a vector multiplet $(A_\mu,\lambda)$ (in our notation $\psi_Q$ and $\lambda$ are 4d Dirac spinors, respectively left- and right-handed). 
Next, we promote the infinitesimal supersymmetry transformation parameter to the one depending  on the 4d spacetime coordinates $x_\mu$.  We also couple the brane fields to the 4d metric induced at the brane, as dictated by general coordinate invariance. 
\item {\it Noether current.} After completing the first step, supersymmetric variations of the brane action no longer  vanish. Firstly, there is a variation of the form  
 \linebreak 
\mbox{$\delta \cl = - \delta(x_{5})\delta(x_{6}) \pa_\mu \eps \, j_{\rm SC}^\mu$},  
where $j_{\rm SC}^\mu$ is the supercurrent of brane matter and gauge  fields.
 In order to cancel this variation  we  need to couple the brane supercurrent to the positive parity gravitino $\psi_\mu^+$.
\item {\it Bianchi identity.} The other uncanceled supersymmetric variations after the first step are those  of the metric fields in the brane action. It turns out that the modifications needed to cancel these variations  can be concisely summarized as a redefinition of  the three-form field strength $H$. That is, we replace the three-form $H$ in the 6d bulk action \erefn{6da} and supersymmetry transformations \erefn{6ds} by $\hat H$ defined such that $d \hat H \neq 0$. With appropriate choice of the right hand side of the Bianchi identity all  variations of the metric at the brane are canceled.  
\item {\it Moduli coupling.} In the final step we  study two-fermion variations containing derivatives of the  dilaton and higher dimensional metric fields. 
From cancellation of these terms we are able  to determine how these moduli should couple to the brane action.  
\ee 

Performing this program we arrive at the bulk-brane action  ($\delta(x_{56}) \equiv
\delta(x_{5})\delta(x_{6})$): 
\begin{align}\label{e.6bba}
  \begin{split}
    \cl &= \cl_{\rm bulk} (H \to \hat H) + \delta(x_{56}) \cl_{\rm brane} \, ,\\ 
    \cl_{\rm brane} &= e_4 \left[ e^{- \Phi} D^\mu Q^\dagger D_\mu Q + {i \over
        2} e^{- \Phi}  \ov \psi_Q \gamma^\mu D_\mu \psi_Q - {i \over 2}  e^{- \Phi} \ov D_\mu
      \psi_Q \gamma^\mu \psi_Q \right.\\
    &\mspace{50mu} - {1 \over 4} F_{\mu\nu}F^{\mu\nu} + i \ov \lambda \gamma^\mu D_\mu \lambda
    - i g \rt  e^{- \Phi} Q \ov \psi_Q \lambda + \hc - {1 \over 2} g^2 e^{- 2 \Phi} |Q|^4 \\ 
    &\mspace{50mu}- {1 \over \rt}  e^{- \Phi} \ov \psi_Q \gamma^\mu  \gamma^\nu  \psi_\mu D_\nu
    Q - {i \over \rt} e^{- \Phi}  \ov \psi_Q \gamma^\mu  \chi D_\mu Q + \hc\\ 
    &\mspace{50mu}\left. - {i \over 4} \ov \lambda \gamma^\mu \gamma^{\nu\rho} \psi_\mu F_{\nu\rho} +
    {1 \over 2} g |Q|^2  e^{- \Phi}  \ov \lambda \gamma^\mu \psi_\mu  - i g |Q|^2  e^{- \Phi}
    \ov \lambda \chi + \hc  \right] \, . 
  \end{split}
\end{align}
In the above formula we used the following definitions
\begin{align}
  \begin{split}
    \hat H_{\mu \dot 5 \dot 6}  &=   
      H_{\mu \dot 5 \dot 6} +  {1 \over M_6^4} \delta(x_{56}) j_\mu \, ,\\
    \hat H_{\tau\rho\sigma}   &=  
      H_{\tau\rho\sigma} + {1 \over M_6^4} \delta(x_{56}) j_{\tau\rho\sigma} \, ,\\
    j_\mu   &= 
      - {i \over \sqrt 2} ( Q^\dagger D_\mu Q - D_\mu Q^\dagger Q) - {1 \over 2 \rt } \ov\psi_Q
      \gamma_\mu \psi_Q  + {1 \over 2 \rt } e^{\Phi} \ov\lambda \gamma_\mu \lambda     \, , \\
    j_{\tau\rho\sigma}  &=  
      - {i \over 2 \sqrt 2}  \ov\psi_Q \gamma_{\tau\rho\sigma}\psi_Q + {1 \over  \sqrt 2}  A
      e^{\Phi} \ov\lambda \gamma_{\tau\rho\sigma}\lambda  \, ,\\
    D_\mu Q &= \pa_\mu Q + i g A_\mu Q \, , \qquad D_\mu \psi_Q = \pa_\mu \psi_Q + i g A_\mu \psi_Q +
    {1 \over 4} \omega_{\mu m n} \gamma^{ m n} \psi_Q \, ,\\
    D_\mu \lambda &= \pa_\mu \lambda + {1 \over 4} \omega_{\mu m n} \gamma^{ m n} \lambda \, .
  \end{split}
\end{align}
The supersymmetry transformations of the bulk fields are those of \eref{6ds} with $H \to \hat H$. 
In addition we need to  modify the transformation law of the Kalb-Ramond form by
\beq
\label{e.6btym} \ds
\delta B_{\dot 5 \dot 6} = \ds  - {i \over 2} {1 \over M_6^4} \delta(x_{56}) 
\ov \psi_Q \eps Q + \hc  \, .
\eeq 
The supersymmetry transformations of the brane fields up to three-fermion terms are simply
\begin{align}\label{e.6gstym}
  \begin{split}
    \delta Q &= {1 \over \rt} \ov \eps \psi_Q \, , \qquad \qquad\qquad  \delta \psi_Q  = - {1 \over \rt} i
    \gamma^\mu D_\mu Q \, \eps \, , \\
    \delta A_\mu &= - {i \over 2} \ov \lambda \gamma_\mu \eps + \hc \, , \qquad\qquad \delta \lambda = {1 \over
      4} \gamma^{\mu\nu} \eps F_{\mu\nu} + {i \over 2} g |Q|^2 e^{- \Phi} \eps \, .
  \end{split}
\end{align}

One comment is in order here. Substituting $H \to \hat H$ in \eref{6bba} results in appearance of singular 
$\delta(x_{56})^2 = \delta(x_{5})^2 \delta(x_{6})^2$
 terms in the bulk-brane action. A similar thing happens in 5d \cite{aa} and in that case  the situation is well understood. 
 Firstly, these singular terms are absent in the low energy effective theory after integrating out the Kaluza-Klein modes. Secondly, in the 5d set-up the $\delta^2$ terms provide for necessary  counterterms to cancel divergences in certain one-loop diagrams \cite{mipe}. In 6d  the latter issue was studied in \cite{le} with the same conclusion. 
 We will show in \sref{lees} that the former is also true in 6d:  the $\delta(x_{56})^2$ terms cancel  after compactification to 4d and integrating out the Kaluza-Klein modes of the Kalb-Ramond form. 

We can further generalize the setup by allowing for  a superpotential $W$ for the brane  chiral multiplet. Here we give the supersymmetric bulk-brane action in the simplest case when  
$W = W_0 = {\rm constant}$. In other words, we supersymmetrize a gravitino brane mass term. The brane action in addition to \erefn{6da} should contain
\beq
\label{e.6bgm}
\cl_{w} =  \frac{1}{2}e_4 \delta(x_{56}) \sqrt{A \over \tau_2 } W_0  \left(
-\ov \psi_\mu \gamma^{\mu\nu} C \ov\psi_\nu^T 
 + i  \ov \psi_{5+6} \gamma^{\mu} C \ov {\psi_\mu}^T
 + i  \ov \psi_{5-6} \gamma^{\mu} C \ov {\psi_\mu}^T 
+ \hc \right) \, ,
\eeq
where $\psi_{5\pm 6} = - (\psi_{5} \pm i \psi_{6})$. 
Moreover, we need to modify the transformation laws of the higher dimensional components of the gravitino as 
\bea
\label{e.6bgms}
\delta \psi_{5\pm 6} &=& \pm {1 \over M_6^4}\delta(x_{56})  
{1 \over \sqrt{A \tau_2}} W_0   C \ov {\eps}^T 
\eea 
In the above $A$ and $\tau_2$ are related to  the volume and the shape of the compact torus. Their definition in terms of   higher dimensional components of the 6d metric 
reads: $A/\tau_2 = g_{\dot 5 \dot 5}$, $A^2 = {\rm det} g_{\alpha\beta}$.   

One very important conclusion is that the couplings of the bulk moduli to the brane are
completely fixed and non-trivial. Namely, we found that the dilaton must couple to the kinetic
terms of the brane chiral multiplet  via the exponential factor $e^{-\Phi}$. On the other hand,
the dilaton does not couple  to the kinetic terms of the brane gauge fields. In the presence of
a brane superpotential also the volume and the shape of the compact torus must couple to the
brane. One interesting  consequence follows. If some brane dynamics develops a vev of an
$F$-term and/or $D$-term potential   
then the bulk moduli couple to the resulting brane tension as 
\beq
\lb{e.fdv}
\cl = - e_4 \delta(x_{56}) \left ( 
{1 \over 2} e^{- 2 \Phi} g^2 D^2 +  {A e^{\Phi} \over \tau_2} |F|^2   \right ) \, .
\eeq 
These couplings will be important for our discussion  of  moduli stabilization further in this paper. More generally, they should affect the analysis of domain wall solutions in 6d supersymmetric brane-worlds.

\section{Low energy effective supergravity}
\lb{s.lees}

At energies below the compactification scale $M_c$ 
we can integrate out all Kaluza-Klein modes and maintain only the zero modes in the spectrum. 
If the scale of  spontaneous supersymmetry breaking is below $M_c$ the effective theory has to reduce to 4d $N=1$ supergravity. In this section we identify the \kahler potential, superpotential and gauge kinetic function of the effective supergravity.  

We  compactify  the 6d theory on the flat background
\beq
\label{e.fb}
ds^2 =  {1 \over A} g_{\mu \nu} dx^{\mu} dx^{\nu} 
- g_{\alpha \beta} dx^{\alpha} dx^{\beta} 
\eeq
where $\la g_{\mu\nu} \ra = \eta_{\mu\nu}$ and the metric on the two-torus is parametrized as
\beq 
\lb{e.tp}
g_{\alpha \beta} = {A \over \tau_2} \left [ \ba{cc}
1 & \tau_1 \\ 
 \tau_1  &  \tau_1^2+ \tau_2^2 
\ea \right ] \, .
\eeq
The modulus $A$ describes  the volume while  $\tau = \tau_2 + i \tau_1$ the shape of the toroidal extra dimensions. 
  
The light degrees of freedom in the compactified theory are the zero modes (those independent of $x_5,x_6$) of the positive parity fields. They include the metric $g_{\mu\nu}$, the gravitino $\psi_{\mu}^+$, the volume modulus $A$, the shape modulus $\tau$, the dilaton $\Phi$, the Kalb-Ramond form $B_{\mu\nu}$ and 
$B_{\dot 5 \dot 6}$, the dilatino $\chi_+$ and the two extra dimensional components of the gravitino $\psi_5$,  $\psi_6$. 
 Counting the degrees of freedom we find that the zero modes can be collected into  a gravity multiplet and three chiral multiplets (the moduli multiplets). Apart from that there is the brane chiral multiplet $(Q,\psi_Q)$ and the brane gauge multiplet $(A_\mu,\lambda)$.   

{ 
Let us first discuss the relevant scales in our setup.
The input parameters in 6d are $M_6$, $R$ and the moduli vevs $A$ and $\tau$.
In fact, $R$ is a superfluous parameter that can be set to any value by coordinate transformations,
but we find it convenient to keep it explicit.   
For simplicity,  we will assume $\tau \sim 1$, so that the two extra dimensions are of comparable size.
In the 6d Einstein frame the fundamental scale at which gravity gets strong is the 6d Planck mass $M_6$, the KK excitations start at $A^{-1/2} R^{-1}$, and the zero-mode graviton exchange is controlled by $2 \pi R A^{1/2} M_6^2$. 
The 6d field theory description is sensible when the physical volume of the torus $V_{T_2} = (2 \pi R)^2 A$ is large enough,  so that the compactification scale and the cutoff scale are separated by at least an order of magnitude: 
$M_6/(A^{-1/2} R^{-1}) \sim V_{T_2}^{1/2} M_6 \gg 1$. 
This condition also ensures perturbativity of gravitational loop corrections.    
The 4d effective action is formulated in the 4d Einstein frame \erefn{fb}, in which all mass parameters are rescaled by the factor $A^{-1/2}$ wrt the 6d Einstein frame. 
In particular the zero-mode graviton exchange is controlled by the 4d Planck scale $M_p = 2 \pi R M_6^2$, while  the compactification scale is $M_c \sim (A R)^{-1}$.
Of course the ratios of physical scales in both frames are the same and depend only on the physical volume in fundamental units.  
}
\subsection{Tree-level effective action}

Let us  study kinetic terms of the bosonic fields. We obtain 
\begin{align}\lb{e.d2}
  \begin{split}
    \cl_{\rm b \, kin} &= M_p^2 \sqrt{-g} \left[{1 \over 2} R(g) + {1 \over 2 A^2} (\pa_\mu A)^2
      + {1 \over 2} (\pa_{\mu}\Phi)^2 + {1 \over 4 \tau_2^2} (\pa_\mu \tau_2)^2 + {1 \over 4
        \tau_2^2} (\pa_\mu \tau_1)^2 \rule{0pt}{25pt}\right.\\
    &\mspace{100mu}+ {e^{-\Phi} \over M_p^2 A} \pa_\mu Q^\dagger  \pa^\mu Q - {1 \over 4 M_p^2}
    F_{\mu\nu}F^{\mu\nu}+  {e^{- 2 \Phi} \over 12 A^2}  (H_{\mu\nu\rho})^2\\
    &\mspace{100mu}\left.  +{ e^{- 2 \Phi} \over 2
      A^2} \left(\pa_\mu B_{\dot 5 \dot 6} + {i \over \rt M_p^2}(Q^\dagger \pa_\mu Q -\pa_\mu
      Q^\dagger Q) \right)^2 \right]\, ,
  \end{split}
\end{align}
 All the terms except for the last one are trivially obtained by inserting the background  \erefn{fb} into the 6d action \erefn{6bba} and integrating over the extra dimensions. The last term requires a more careful derivation.  The 6d bulk-brane action contains 
\begin{align}\label{e.d3} 
  \begin{split}
    \cl &= {1 \over 12 } e_6 M_6^4 e^{-2 \Phi}\hat H_{A B C} \hat H^{A B C}\\
    &\longrightarrow {e^{-2 \Phi} \over 2 A^2 } \sqrt{-g} M_6^4 \left ( \pa_\mu B_{\dot 5 \dot 6} + \pa_6
      B_{\mu \dot 5} + \pa_5 B_{ \dot 6\mu} + {1 \over M_6^4} \delta(x_{56}) j_\mu   \right )^2 \, ,
  \end{split}
\end{align}
where $j_\mu ={i \over \sqrt 2}( Q^\dagger \pa_\mu Q -\pa_\mu Q^\dagger Q)$ + fermionic terms. 
Compactifying to 4d one cannot simply set the odd components  $B_{\mu \alpha}$ of the Kalb-Ramond form to zero. The reason is  they couple to the brane matter 
via   
$ \delta(x_{56}) \pa_\alpha B_{\mu \beta} j^\mu \to m_n B^{\bn}_{\mu \beta} j^\mu$.
 As the coupling is proportional to the mass, the diagram with an exchange of the massive vector $B^{\bn}_{\mu \beta}$ does not decouple at low energies, and thus integrating out these modes is non-trivial. Restricting to the  tree-level diagrams and  two-derivative terms in the effective action there is a simple  prescription to take into account these non-decoupling diagrams. Namely, one should first solve the equations of motion for $B_{\mu \alpha}(x_5,x_6)$ with $j_\mu$ and $\pa_\mu B_{\dot 5 \dot 6}$ treated as constant sources, then insert the solution back into the 6d action and integrate over the extra dimensions. The non-trivial equations of motion read 
$ \pa^5 \hat H_{\mu \dot 5 \dot 6} = \pa^6 \hat H_{\mu \dot 5 \dot 6} = 0$,
thus 
$ \hat H_{\mu \dot 5 \dot 6} = C_\mu$
where  $C_\mu$ is a constant. Let us choose an ansatz:  
$B_{\mu \dot 5} = - \pa_6 W_\mu$, $B_{\mu \dot 6} =  \pa_5 W_\mu$.
Then the equations of motion reduce to 
\beq
\label{e.d5}
(\pa_5^2 + \pa_6^2)W_\mu = 
\pa_\mu B_{\dot 5 \dot 6}  - C_\mu +  {1 \over  M_6^4} \delta(x_{56}) j_\mu \, .
\eeq
We can solve this equation (with appropriate periodicity conditions) if we choose the integration constant as 
\beq 
\label{e.icc}
C_\mu = \pa_\mu B_{\dot 5 \dot 6} + {1 \over (2\pi R)^2 M_6^4} j_{\mu} \, .
\eeq 
This follows from integrating  the equation \eref{d5} over the torus and noticing that the left-hand side vanishes by the Stokes theorem. 
The equation of motion can now be solved \cite{lenizu}, but this is not important for our purpose. Inserting  $\hat H_{\mu \dot 5 \dot 6} = C_\mu$  back into the 6d action  and integrating over the extra dimensions we  reproduce the last term in  \eref{d2}. 

The kinetic terms \erefn{d2} are  not diagonal yet as both $A$ and $\Phi$ couple to the axions 
$B_{\dot 5 \dot 6}$ and $H_{\mu\nu\rho}$. Therefore we define new variables \cite{quevedo}:
\beq
s = A  e^{- \Phi} \, , \qquad t = A  e^{ \Phi}  \, .
\eeq 
We also dualize the two-form to a scalar:  
$e^{-2 \Phi} H_{\mu\nu\rho} = \eps_{\mu\nu\rho\tau} \pa^\tau \sigma$.
Then the kinetic terms take the form 
\begin{align}\label{e.boskin}
  \begin{split}
    \cl_{\rm b \, kin} &= M_p^2 \sqrt{-g} \left[{1 \over 2} R(g) + {1 \over 4 t^2} (\pa_\mu
      t)^2  + {1 \over 4 s^2} (\pa_{\mu} s)^2+ {1 \over 4 \tau_2^2}(\pa_\mu \tau_2)^2 +  {1
        \over 4 \tau_2^2} (\pa_\mu \tau_1)^2 \rule{0pt}{25pt}\right.\\
    &\mspace{98mu}+ {1 \over M_p^2 t} \pa_\mu Q^\dagger  \pa^\mu Q - {1 \over 4 M_p^2}
    F_{\mu\nu}F^{\mu\nu} \\
    &\mspace{98mu}\left. + {1 \over 2 s^2}(\pa_\mu \sigma)^2  + {1 \over 2 t^2} \left (\pa_\mu B_{\dot
          5 \dot 6} + {i \over \rt M_p^2}(Q^\dagger \pa_\mu Q -\pa_\mu Q^\dagger Q) \right )^2
    \right] \, . 
  \end{split}
\end{align}

These kinetic terms are reproduced by the \kahler potential and the gauge kinetic function:
\bea & \ds
\label{e.bbk}
K = - \log \left ({1 \over 2}(T + \ov T) - {1 \over M_p^2} |Q|^2 \right ) 
- \log \left ({1 \over 2}(S + \ov S) \right )
 - \log \left ( {1 \over 2}(\tau + \ov \tau) \right ) \, ,
\nl
f = 1 \, ,
\eea
with the following definition of the scalar components of the moduli multiplets: 
\beq
\label{e.bbkf}
  T  =    t +  {1 \over  M_p^2} |Q|^2 +  i \rt B_{\dot 5 \dot 6}  \, , \qquad
  S  =    s + i \rt \sigma \, , \qquad
  \tau   =    \tau_2 + i \tau_1 \, .
\eeq
Note that the admixture of the brane matter in the $T$-modulus is necessary to correctly reproduce the $|Q|^2 (\pa Q)^2$ terms in \eref{boskin}.   

When the brane superpotential is present, compactification on the background of \eref{fb}
yields the gravitino mass term%
\footnote{This formula is true at lowest order in $W_0$. At higher orders in $W_0$ the mass of the gravitino zero mode receives divergent corrections that require tree-level renormalization \cite{gowi}. This UV sensitivity of the effective theory can be neglected  when 
$W_0/M_p^2$ is much smaller than the compactification scale, which we always assume in this paper.}  
\beq
\lb{e.4gmt}
\cl = - {1 \over 2 A \sqrt{\tau_2} } \sqrt{-g} W_0
\ov \psi_\mu \gamma^{\mu\nu} C \ov\psi_\nu^T + \hc  \, .
\eeq 
Comparing it with the gravitino mass term in 4d \sugra,  
$ - {1 \over 2} \sqrt{-g} e^{K/2} W$$\ov \psi_\mu \gamma^{\mu\nu} C \ov\psi_\nu^T$
 and the \kahler potential \erefn{bbk} we find that \erefn{4gmt} corresponds  to a constant superpotential:
\beq
\label{e.w0}
W = W_0  \, .
\eeq

We can now easily generalize these results to the case when brane matter with arbitrary kinetic terms  is present on  all four $\mathbb{Z}_2$ fixed points.  The low energy effective supergravity is then described by 
\begin{align}\label{e.bbkg}
  \begin{split}
    K &= - \log \left ({T + \ov T\over 2}  -  \Omega_{\rm branes} \right ) - \log \left ( {S +
      \ov S \over 2} \right ) - \log \left ({\tau + \ov \tau \over 2} \right ) \, , \\
    \Omega_{\rm branes} &=  \sum_a \Omega_{a} (Q_{i}^a,Q_{i}^a{}^\dagger)  \, ,
\qquad f =  \sum_a f_{a}(Q_i^a) \, ,
\qquad W = \sum_a  W_{a}(Q_i^a) \, ,
\end{split}
\end{align}
where $a = 1 \dots 4$ labels the $\mathbb{Z}_2$ fixed points.  One also needs to generalize the definition of $T$ as 
\beq
\lb{e.bbkfg}
\re T  =   t +  \Omega_{\rm branes} \, .
\eeq
The effective supergravity defined by \eref{bbkg} is the starting point for our study of supersymmetry breaking mediation, which we perform in \sref{gmsb}.   
Here we point out several interesting features of \erefn{bbkg}: 
\be
\item  { The \kahler potential \eref{bbkg} is not of the sequestered form.
Still, the higher dimensional locality  is encoded in the fact that 
$\Omega_{\rm branes}$, $f$ and $W$ consist of separate contributions from different branes.
 One consequence of this special structure is that the 4d effective action derived from \erefn{bbkg} does not contain `non-local' interactions that would couple scalar operators involving fields from different branes. 
On the other hand, there are `non-local' vector-vector interactions in 
$\cl \sim (\pa_i \Omega \pa_\mu Q_i)^2$ 
that arise from integrating out the KK modes of the bulk 2-form field.
}
\item {The bulk moduli $T$, $S$ and $\tau$ couple universally to all brane fields. 
In our setup there is no possibility to introduce non-universal couplings of these moduli, as the form of \eref{bbkg} is uniquely fixed by 6d local supersymmetry. 
Therefore soft terms mediated by these moduli will necessarily be flavour blind. }
\item  The superpotential $W$ does not  depend on the moduli $T$, $S$, $\tau$. This statement remains true as long as only perturbative effects are considered. Later in \sref{ms} we will argue that non-perturbative bulk physics (e.g. gaugino condensation) can generate moduli dependence of the superpotential. 
\item  The \kahler potential is of the special no-scale form 
such that the coefficients of the logs sum up to three  and therefore  the factor $-3 |W|^2$ in the scalar potential cancels out.
In consequence, for $W$ independent of the moduli, the scalar potential takes the form
\beq
V = {1 \over s \tau_2} g^{i \ov j} \pa_i W \ov{\pa_j W} \, ,
\eeq   
where the indices $i,j$ run over  brane matter fields only and $g^{i \ov j}$ is the inverse sigma-model metric in the brane sector. 
\item  The parametrization \erefn{tp} of the compact manifold is redundant, so that the shape moduli related by a $SL(2,\ZZ)$ transformation $\tau \to {a \tau -  i b \over c i \tau +  d}$ with integer $a,b,c,d$ and $a d - bc = 1$ parametrize the same torus. 
 Under the two generators of  $SL(2,\ZZ)$ transformations 
 $g_1:\tau \to \tau + i$   and  $g_2: \tau \to 1/\tau$ 
the \kahler potential \erefn{bbkg} transforms as
\beq
g_1:\ K \to K \, , \qquad \qquad 
g_2:\ K \to K + \log \tau + \log \ov \tau \, .
\eeq
This ensures that  the kinetic terms are invariant under the  $SL(2,\ZZ)$ transformations. 
 However if the superpotential is non-vanishing it should   transform as
$ g_1:\ W \to W$, $g_2:\ W \to {1 \over \tau} W$ in order for the entire supergravity action to be invariant. Since the superpotential in \erefn{bbkg} is independent of $\tau$   its presence breaks the $g_2$ generator of  $SL(2,\ZZ)$.  
Similarly, the presence of vector fields on the branes breaks the $SL(2,\ZZ)$  invariance.  
\item It may be surprising that the structure of the \kahler potential \erefn{bbkg} is so different than that of the \kahler potential \erefn{o5t0} derived from 5d minimal supergravity. Note however that compactifying our 6d model to 5d we obtain a 5d supergravity coupled to a tensor and a vector multiplet. Therefore there is no a priori reason why the minimal 6d model should be similar to the minimal 5d model.   
\ee

\subsection{One-loop corrections}

Integrating out heavy Kaluza-Klein modes at one-loop order corrects the \kahler potential of the low-energy effective supergravity given in \eref{bbkg}.
{ In 5d the minimal supergravity model yields  the no-scale tree-level \kahler potential, therefore one loop corrections are crucial for a discussion of supersymmetry breaking \cite{ghri,bugago,rascst}. In 6d the no-scale structure is absent and generically we do not expect loop corrections to play a prominent role. 
However we will see in \sref{ms} that in certain circumstances tree-level mediation of supersymmetry breaking may be absent also in 6d. In such case it is important to know the precise form of one loop corrections. 
Our result for the one-loop \kahler potential may also be interesting for future applications, as e.g. it encodes the Casimir energy contribution to the scalar potential. }

 As discussed in refs. \cite{rascst,aa} there is a simple formula for the one loop \kahler potential:
\beq  
\label{e.do}
\Delta \Omega = { N \over 3} {\Gamma(1-d/2)\over M_p^2 (4 \pi)^{d/2}} \, .
\sum m_n^{d-2}
\eeq
where 
$K = -3 \log (\Omega_{\rm tree} + \Delta \Omega)$, $d = 4 +\eps$ and $m_n$ denotes  the Kaluza-Klein spectrum in the 4d conformal frame. The numerical factor $N$ can be determined by counting the number of fermions and gauge bosons (from the 4d perspective) at each Kaluza-Klein level:
$N = N_{1/2} - 2 N_{1}$.  
In the present case we obtain $N = - 4$. 
The spectrum in the conformal frame is given by: 
\beq 
m_{n_5,n_6}^2 = {1 \over A^{4/3} \tau_2^{2/3} R^2} |n_6 + i n_5 \tau|^2
\eeq
Using the standard methods \cite{popo} we find the sum:
\begin{align}
  \begin{split}
    \Sigma(s) &\equiv {1 \over 2}\sum'_{n_5,n_6\in Z}m_{n_5,n_6}^{2 s} = {1 \over A^{4/3}
      \tau_2^{2/3} R^2} \left(\Sigma_1(s)+\Sigma_2(s) + \Sigma_3(s) \right), \\
    \Sigma_1(s) & =  \ds \zeta(-2 s),\mspace{120mu} \Sigma_2(s)  = \ds \sqrt{\pi} \tau_2^{2s+1}
    {\Gamma(-s-1/2) \zeta (-2 s -1) \over \Gamma(-s)},\\
    \Sigma_3(s) & = {2 \tau_2^{s+1/2} \over \pi^s \Gamma(-s)}\sum_{n_5,p_6>0}  \left(\left
        ({n_5 \over p_6} \right )^{s+1/2} K_{s+1/2}(2 \pi \tau_2 n_5 p_6) e^{ - 2\pi i\tau_1
        n_5 p_6} + \hc \right ).
  \end{split}
\end{align}
Inserting $s = d/2 -1$ and then $d = 4 + \eps$ we find that all $\Sigma_i$'s are proportional to $\eps$, which cancels 
the $1/\eps$ pole of $\Gamma(1-d/2)$ in \eref{do}. Thus our expression for the  correction to the \kahler potential is finite and we find:
\begin{align}\lb{e.6do}
  \begin{split}
    \Delta \Omega &=  - {32 \over 3 (4 \pi)^2} {1 \over (2 \pi R M_p)^2} \left (T + \ov T -  {2
        \over M_p^2} \Omega_{\rm branes} \right)^{-2/3} \left (S + \ov S \right)^{-2/3}\left
      (\tau + \ov \tau\right)^{-2/3}  \\
    &\quad \mspace{5mu} \times \left[\zeta(3) + {\pi^3 (\tau + \ov \tau)^3 \over 360} + \sum_{n_5,p_6>0} {1
        + \pi n_5 p_6(\tau + \ov \tau) \over p_6^3}( e^{ - 2\pi \tau n_5 p_6} + \hc )  \right]
  \end{split}
\end{align}
{ For $\tau \sim 1$ this is suppressed wrt to the tree level term by the factor 
${1 \over (4 \pi)^2 (2 \pi R M_p A)^2}$ or, equivalently,  
${1 \over (4 \pi)^2 (V_{T_2} M_6)^{4}}$.
The volume suppression signals that $\Delta \Omega$ corresponds to (truly) non-local operators in 6d.}
Apart from the finite part \erefn{6do} there is a UV sensitive contribution to $\Delta \Omega$ but it simply renormalizes the tree-level parameters: the Planck scale and $\Omega_{\rm branes}$ in  the \kahler potential. 
One can also compute one-loop gravitational corrections to the gauge kinetic function $f$ and  find that they vanish (except for a  UV sensitive  renormalization of the tree-level~$f$).

\section{Gravity mediated supersymmetry breaking} 
\lb{s.gmsb}

We have gathered all necessary tools to analyze supersymmetry breaking in 6d brane-worlds. 
This issue can be  most efficiently studied within an  effective theory after all Kaluza-Klein modes are integrated out, that is within 4d supergravity defined by \eref{bbkg}. 
We will see that, unlike the tree-level \kahler potential in the 5d case, 
the \kahler potential \erefn{bbkg} may lead to gravity mediation of supersymmetry breaking. 
In order to recognize what is special about the 6d case let us first review the corresponding analysis in 5d \cite{lusu,rascst}.  
For the minimal flat 5d brane-world supergravity the tree-level \kahler potential of the low energy 4d  supergravity reads:
\beq
\label{e.5ot}
\Omega_{5d} = {1 \over 2} (T + \ov T)
 - {1 \over 3}\Omega_V(Q_V,Q_V^\dagger) 
 - {1 \over 3}\Omega_H(Q_H,Q_H^\dagger) \, .
\eeq
Here $Q_V$ are superfields living on the visible brane, while $Q_H$ correspond to the hidden brane sector. We also expand the visible matter \kahler potential as 
$\Omega_V = |Q_V|^2/M_p^2 + \dots$.  
The visible matter couples neither to the hidden brane matter nor to the moduli  in \eref{5ot}. This so-called  sequestering guarantees the absence of  tree-level gravity mediation of supersymmetry breaking. This is most easily seen in the conformal compensator formalism in which  the relevant part of the action can be written as 
$\cl =  -3 M_p^2 \int d^4 \theta  C^\dagger C \Omega$,  
with $C = 1 + \theta^2 F_C$. 
Equivalently, we can study  the soft scalar masses directly in the on-shell formulation. The standard expression for the scalar potential in  4d supergravity in the  Einstein frame  reads 
\cite{crjusc} 
\beq 
V= {1 \over M_p^2} e^K (K^{i \ov j} D_i W \ov{D_j W} - 3 |W|^2) 
+ {M_p^4 \over 2\re f_a} D_a^2 \, .
\eeq
The scalar masses are defined as
\beq
\label{e.mqsg}
m_{Q_V}^2 = \left . 
{ {\pa^2 V \over \pa Q_V \pa Q_V^\dagger} \over  {\pa^2 K \over \pa Q_V \pa Q_V^\dagger} }
\right |_{Q_V=0} \, .
\eeq
The soft masses can be expressed in terms  of the supersymmetry breaking order parameters, 
the $F$-terms and the gravitino mass defined by:
\beq
\label{e.ftd}
F_i = - e^{K/2} (K^{-1})_{i j} D_j W \, ,
\qquad \qquad
m_{3/2} = {1 \over M_p^2} e^{K/2} W  \, ,
\eeq 
where $D_j W = \pa_j W + \pa_j K \, W$ and $K$ and $W$ should be  evaluated for   moduli vevs at the minimum. Tuning the cosmological constant to zero allows to relate $m_{3/2}$ and $F_i$, therefore the soft masses can be expressed solely by $F$-terms (and $D$-terms if present).  

Performing this procedure in the 5d setup we find at tree-level
\beq
m_{Q_V}^2 = 0 
\eeq 
 for arbitrary  values of $F_T$, $F_H$ and $m_{3/2}$. Independence of the soft scalar masses of supersymmetry breaking order parameters is the consequence of the no-scale structure of \eref{5ot}.   
At one-loop level the \kahler potential \erefn{5ot} gets corrected by \cite{rascst,aa} :
 \bea & 
\label{e.5ol}\ds 
\Delta \Omega_{5d} =  \eps (T + \ov T - {2 \over 3}\Omega_V  - {2 \over 3 }\Omega_H)^{-2} 
\nl \approx
\eps \left [  (T + \ov T)^{-2} + \frac{4}{ 3M_p^2} (T + \ov T)^{-3} |Q_V|^2  
+  \frac{8}{3 M_p^4} (T + \ov T)^{-4} |Q_V|^2 |Q_H|^2 + \dots
\right ] \, , 
\eea 
where  $\eps = - {16 \zeta(3) \over 3 (4 \pi)^2} {1 \over (2 \pi R M_p)^2}$.
We see that the sequesterd structure of \eref{5ot} is not respected by  loop corrections. 
One finds (here and in the following we assume that field vevs in the hidden brane sector are negligible):  
\beq
m_{Q_0}^2 = 48 \eps {|F_T|^2 \over (T + \ov T)^5} +  8 \eps {|F_{H}|^2 \over M_p^2 (T + \ov T)^4} \, , \eeq
 so at one-loop level both radion and brane-to-brane mediation are present. But, since $\eps$ is negative in the minimal 5d  set-up,  this contribution to scalar soft masses squared is negative and thus gravity cannot be the leading source of  supersymmetry breaking mediation. Note also that the anomaly mediation, which is another mechanism operating here,  by itself cannot save the day  as it  generates negative slepton masses.

We move to the 6d case. For simplicity we consider matter only on two of the four fixed points.
 At  tree-level: 
\beq
\label{e.6ot}
\Omega_{6d} = {1 \over 2}
\left (T + \ov T - 2 \Omega_V  - 2 \Omega_H \right)^{1/3}
\left (S + \ov S\right)^{1/3}  \left (\tau + \ov \tau\right)^{1/3} \, .
\eeq
It is clear that sequestering is not present in the \kahler potential \erefn{6ot}.
 This is confirmed by a direct computation of the scalar soft masses: 
\begin{align}\lb{e.6sm}
  \begin{split}
    m_{Q_V}^2 &= {1 \over 3} {|F_S|^2 \over (S + \ov S)^2} + {1 \over 3} {|F_\tau|^2 \over
      (\tau + \ov \tau)^2}- {2 \over 3} {|F_T|^2 \over (T + \ov T)^2} - {4 \over 3} {|F_H|^2
      \over M_p^2 (T + \ov T)} \, , \\
    m_{3/2}^2  &= {1 \over 3} \left({|F_T|^2 \over (T + \ov T)^2}  +  {|F_S|^2 \over (S + \ov
        S)^2} + {|F_\tau|^2 \over (\tau + \ov \tau)^2} + 2 {|F_H|^2 \over M_p^2(T + \ov T)} \right )
    \, . 
  \end{split}
\end{align}

Thus in 6d tree-level gravity mediation can be present. 
This result might be a little surprising, as the effective action we derived in \sref{lees} does not contain any interactions that could mediate supersymmetry breaking. 
However, the analysis in  \sref{lees} did not include possible non-perturbative effects.
In absence of non-perturbative effects the effective superpotential cannot depend on the bulk moduli and the resulting action has a special structure that reflects the higher-dimensional locality. 
Actually, gravity mediation is absent as long as the superpotential is $T$ independent.   
The \kahler potential depends on $T$ and the brane superfields $Q$ only via the combination 
$T + \ov T - 2 \Omega_{\rm branes}$. 
For $\pa_T W = \pa_Q W = 0$, $T$ and  $Q$ enter the scalar potential $V$ only in this combination. 
Once the potential stabilizes $T$, that is $\pa_T V = 0$ at the minimum, 
for $\Omega_{\rm branes} \sim Q^\dagger Q + \dots$ we find 
\beq
m_{Q}^2  \sim \pa_{Q} \pa_{Q^\dagger } V  \sim  \pa_T V   = 0 \, .
\eeq 
For $\pa_Q W \neq 0$ we could, in general, obtain $m_{Q}^2 \neq 0$, but in such case the mass terms would be supersymmetric. 
We conclude that gravity mediated supersymmetry breaking does not operate as long as $\pa_T W =0$ and moduli are stabilized.
This situation is in fact similar as in the 5d set-up (without large gravity brane kinetic terms). 
In that case, the $T$ and $Q$ enter the \kahler potential in an analogous combination 
(see eqs. \eref{5ot} and \eref{5ol}) and soft masses appear (at one loop) only for $\pa_T W \neq 0$.

In the 5d case gaugino condensation in the bulk induces a $T$ dependent superpotential. 
As we discuss in the next section, in the 6d case a $T$ dependent superpotential in the  effective 4d supergravity can arise only after including in the 6d action certain higher order operators, whose presence can be motivated by cancellation of gravitational anomalies in a fully consistent set-up.
Only in these very special circumstances the 4d effective action may contain gravitational interactions that mediate supersymmetry breaking. In such setup, depending on the relation between the $F$-terms,  the tree-level mediated soft masses can be either positive or negative.

In conclusion, the 4d effective supergravity derived from the 6d minimal model has different property than that derived from 5d. 
Most importantly, the \kahler potential is not of the sequestered form. 
In spite of that the action does not lead to gravity mediation at tree-level, unless 6d action is augmented by certain higher order operators and non-perturbative effects introduce $T$ dependence in the effective superpotential.

We close this section with a more general formula for scalar soft mass terms in the presence
of  non-zero $D$ -term vev on the hidden brane. From  \eref{fdv}, the $D$-term vev contributes to the action as   $\cl =- {1 \over 2} e_4 \delta(x_{56}) e^{-2 \Phi} D^2$.  After compactification to 4d on the background \erefn{fb} we obtain:
 \beq
\label{e.6lp}
V_{D} =  {1 \over 2} D^2 \left [ (T + \ov T)/2
 -  \Omega_V  -   \Omega_H  \right ]^{-2}
\eeq 
When tuning the cosmological constant to zero it is convenient to eliminate $D^2$. 
Then we obtain:
\begin{align}\label{e.6smd}
  \begin{split}
    m_{Q_V}^2 &=  4 m_{3/2}^2 - 2 {|F_T|^2 \over (T + \ov T)^2}-  {|F_S|^2 \over (S + \ov S)^2}-
    {|F_\tau|^2 \over (\tau + \ov \tau)^2}- 4 {|F_H|^2 \over M_p^2 (T + \ov T)} \, , \\
    {D^2 \over 2 M_p^2 (T/2 + \ov T/2)^2} &=\left(3 m_{3/2}^2 -  {|F_T|^2 \over (T + \ov T)^2} -
      {|F_S|^2 \over (S + \ov S)^2} -   {|F_\tau|^2 \over (\tau + \ov \tau)^2} - 2 {|F_H|^2
        \over M_p^2 (T + \ov T)}\right)  .
  \end{split}
\end{align}
For $D = 0$ we can express $m_{3/2}$ by the $F$-terms and we recover the previous formula~\erefn{6sm}.

\section{Moduli stabilization}
\lb{s.ms}

In this section we discuss possible ways of stabilizing moduli in the 6d set-up. This certainly requires going  beyond the minimal model we constructed in \sref{bba}. Indeed, in the model with the \kahler potential and the moduli independent superpotential of \eref{bbkg},  the scalar potential for the moduli, schematically, is given by
\beq
\lb{e.4fdv}
V = {F^2 \over (S/2 + \ov S/2)(\tau/2 + \ov \tau/2)} +  {D^2 \over 2 (T/2 + \ov T/2)^2} \, ,  
\eeq
which is of the runaway type.      
To conjure up a stabilization mechanism we have to modify the bulk field content. The simplest way is to add non-abelian vector multiplets in the bulk. Kinetic terms for a bulk  gauge field are  of the form 
$\cl = - {1 \over 4} e_6 e^{-\Phi} F_{A B}F^{A B}$ \cite{nise,sase}.  
After compactification on the background \erefn{fb} it becomes  
$\cl = - {1 \over 4} \sqrt{-g} A  e^{-\Phi} F_{\mu \nu}F^{\mu\nu} = - {1 \over 4} S  F_{\mu \nu}F^{\mu\nu}$. 
Therefore a bulk vector multiplet corresponds at low energies to a vector multiplets with the gauge kinetic function $f = S$. Furthermore, one loop corrections yield \cite{ghgr}
$\Delta f = {b \over 4 \pi^2} \log \eta(i \tau)$, where $b$ is the beta function coefficient and
$\eta$ denotes the Dedekind function.
Gaugino condensation in this sector can be described in 4d \sugra by the effective superpotential
$W_{\rm eff} = \Lambda \eta(i \tau)^{-2} e^{- {8 \pi^2 \over b}  S}$.
Note that the presence of the Dedekind $\eta$ function is necessary to render the superpotential the correct transformation under $SL(2,\ZZ)$ invariance  $\tau \to \tau + i$ and $\tau \to 1/\tau$
 (bulk dynamics, as opposed to that of branes, respects the $SL(2,\ZZ)$ invariance). 
We assume that  there are two  condensing bulk gauge groups and that a vev of the brane superpotential is negligible.  Then the superpotential is of the racetrack \cite{racetrack} form:
\beq
\lb{e.rtsp}
W = \eta(i \tau)^{-2} \left ( \Lambda_1 e^{- a_1 S} -  \Lambda_2 e^{- a_2 S} \right ) \, .
\eeq  
A model with such superpotential and the \kahler potential \eref{6ot} stabilizes $S$, $\tau$ 
at an ${\rm AdS}_4$ stationary point. The equations of motion $\pa_S V = 0$, $\pa_\tau V =0$ are satisfied  for $S$ and $\tau$ solving  $F_S = 0$ and $F_\tau =0$. 
We find that $F_S = 0$ for $S$ solving the constraint 
\beq
\lb{e.srt}
\Lambda_1 e^{-a_1 S} (1 + a_1 (S + \ov S)) = 
\Lambda_2 e^{-a_2 S} (1 + a_2 (S + \ov S))  \, .
\eeq 
This fixes both $\re S$ and $\im S$. As is well known, for the racetrack model the  beta functions of the two condensing groups have to be similar,  
$|a_1 - a_2| \ll a_1$, in order for  the solution to \eref{srt} occur for large $\re S$ where the expression \erefn{rtsp} for the superpotential is reliable. 
$F_\tau = 0$ is solved for $\tau$ satisfying:
\beq
\hat G_2(\tau, \ov \tau) \equiv 
-2 \pi \left ( \pa_\tau \log \eta(i \tau) + {1 \over \tau + \ov \tau} \right )
= 0 \, .
\eeq 
Here $\hat G_2$ is the modified Eisenstein series. It has zeros in the fundamental $\tau$ domain at the self dual fixed points 
$\tau = 1$ and $\tau = \sqrt{3}/2 + i/2$. It turns out that the latter solution is a saddle point of the potential while the former is a minimum (in the $S$, $\tau$ plane; in the full potential $T$ is for the time being 
a runaway direction).

We now discuss mechanisms of stabilizing $T$.  
One possibility,  that relies solely on perturbative effect,  
is to assume that some dynamics has generated vevs of $D$-terms and $F$-terms of hidden brane fields, which contributes to the moduli potential as in \erefn{4fdv}.    
 In the presence of such hidden brane vevs the vacuum values of $S$ and $\tau$ are slightly shifted, so that $F_S$ and $F_\tau$ no longer vanish at the true minimum. 
After stabilizing $S$ and $\tau$ the potential for $T$ is of the form:
\beq
V/M_p^2 = -  {2 \hat m_{3/2}^2 - |\hat F_S|^2 - |\hat F_\tau|^2  \over (T + \ov T)} 
+ {2 \over M_p^2} |\hat F_H|^2  + {2 D^2 \over (T + \ov T)^2}
\eeq 
where 
$(\hat m_{3/2},\hat F_S, \hat F_\tau) 
= (m_{3/2},F_S/(S+\ov S),F_\tau/(\tau+\ov \tau))(T + \ov T)^{1/2}$ and 
$\hat F_H =  F_H (T + \ov T)^{-1/2}$
are $T$-independent. Then the solution to $\pa_{T} V = 0$ with $V=0$ at the minimum requires
\beq
\lb{e.st1}
{2 D^2 \over (T + \ov T)^2} = 
 {2 \over M_p^2} |\hat F_H|^2 = 
{\hat m_{3/2}^2 - {1 \over 2}|\hat F_S|^2 - {1 \over 2} |\hat F_\tau|^2  \over (T + \ov T)} 
\eeq   
One of these equations determines $(T + \ov T)$ and the other is a fine-tuning needed to arrive at $V=0$  at the minimum. This solution is always a local minimum for $(T + \ov T)$.

There are, however, two problems with this model of $T$ stabilization. Firstly, the axion  $(T - \ov T)$ is not stabilized. This could be   solved if the axion transformed under some gauge symmetry   analogously as in ref. \cite{quevedo}. In such case the axion would be eaten by the gauge field.  
Secondly, since the superpotential in this model is $T$ independent the soft masses vanish by the argument of the preceding section. Note that this remains true after taking into account the one-loop corrections \erefn{6do}.%
\footnote{
 It was argued  in ref. \cite{aa} that there is an ambiguity in the one-loop \kahler potential at  order  $ \epsilon \Omega_{\rm branes}^3$, depending on  regularization of the bulk fields behavior near the delta-like branes. 
If $\Omega_{\rm branes}$ contains a constant (corresponding to brane field vevs or a brane gravity kinetic term), 
$\Omega_{\rm branes} = L + |Q_V|^2/M_p^2 + \dots$,  
then this  ambiguity would result in appearance of regularization dependent soft masses of order $L^2 F^2$.
This ambiguity can become relevant only when brane vevs or coefficients of  brane gravity kinetic terms are bigger than the compactification scale. 
See also refs. \cite{rascst,grrasc}, which argue that the one-loop computation is non-ambiguous.}   
%
Therefore in this model  gravity mediation cannot be the leading source of supersymmetry breaking in the visible sector and one needs to rely on gaugino mediation \cite{gaugino} or Scherk-Schwarz breaking \cite{le}.   

Another mechanism of $T$ stabilization relies on including 6d anomaly corrections. As discussed in ref. \cite{quevedo} the Green-Schwarz mechanism of anomaly cancellation in 6d requires extending the 6d supergravity action \erefn{6da} by certain higher dimensional operators, including the coupling of the form 
$\cl \sim \alpha  B \wedge F \wedge F$.  At low energies the effect of such coupling can be described by an order $\alpha$ correction to the  gauge kinetic function, $f = S + \alpha T$.  When  gauginos condense,  the effective superpotential picks up   $T$ dependence. We find that in certain multiple condensates models $T$ is indeed stabilized, with $F_T$, $F_S$ and $F_\tau$ much smaller than $m_{3/2}$. The moduli stabilization yields a negative contribution of order $-3 M_p^2 m_{3/2}^2$ to the scalar potential that must be canceled by a contribution from  the brane sector. 
 In absence of hidden brane $D$-term vevs we need $F_H$ of order $m_{3/2}$. In such case, by \eref{6sm} the gravity mediated soft masses are negative. However if the hidden brane $D$-term vev  is of order $m_{3/2}$ and $F_H$ is negligible  then by \eref{6smd} we arrive at positive scalar soft masses squared. 
Furthermore, gaugino masses of similar order can be generated by anomaly mediation.
Although anomaly mediated gaugino masses are suppressed by a loop factor with respect to the gravitino mass, they can be comparable to the tree-level mediated scalar masses.
The reason is that, typically,  non-perturbative moduli stabilization   results in suppression of the corresponding F-terms with respect to the gravitino mass, and the suppression factor can be comparable to the loop factor 
$\sim 4 \pi^2$.
In such case the soft terms  acquire comparable contribution from anomaly and gravity mediation, which leads to interesting phenomenology. 
In fact, a similar pattern of soft terms was recently analyzed (in the framework of KKLT compactification \cite{kakali}) in refs. \cite{chfani,chjeok}.

\section{Conclusions}
\lb{s.c}

In this paper we studied gravity mediated supersymmetry breaking in the 4d effective theory derived from 6d brane-world supergravity compactified  on $T_2/\mathbb{Z}_2$. The main motivation for invoking extra dimensions is that they offer an  opportunity of constructing a viable and predictable theory of contact terms between the visible and hidden sectors. Realizations of this idea in the settings of 5d brane-world supergravity have proven unsatisfactory. { The 5d models yield negative soft scalar  masses squared, unless they are augmented with large gravity brane kinetic terms}. Therefore it is interesting to go beyond five dimensions. 

Using the Noether method we constructed a locally supersymmetric action for a bulk-brane system consisting of the minimal $N=2$ 6d supergravity and $N=1$ vector and chiral multiplets on codimension-two branes located at the orbifold fixed points.
Then we derived an effective action describing physics  at energies below the compactification scale. 
As the coupling of the bulk to the branes turn out to be  uniquely fixed by local supersymmetry, the moduli dependence of the low energy effective  action (and hence the contact terms)  is also fixed. 

The \kahler potential of the 4d low energy effective theory is not of the sequestered form. 
As the consequence, contact terms between the visible and hidden sectors can be present already at tree-level.
The soft terms mediated by the bulk moduli originating from the 6d gravity+tensor multiplet are necessarily flavour blind.   
We find however that, typically, the 4d effective action exhibits similar sequestering properties as that derived from minimal 5d models.
More precisely, in 6d tree-level gravity mediation does not occur as long as the effective superpotential is $T$ independent.  
In order to generate a $T$ dependent superpotential one needs to invoke, simultaneously, higher order operators and  non-perturbative physics.
Turning to such models, we identified one scenario in which all moduli are stabilized at a vacuum with the vanishing cosmological constant and the gravity mediated soft scalar masses squared are positive. This is an example of a successful embedding of 4d supergravity into a more fundamental set-up such that  phenomenologically viable contact terms can be predicted. In this scenario stabilization of the $T$ modulus requires  rather contrived physics (multiple gaugino condensates). We believe however that generalizing the 6d bulk-brane set-up (e.g. to flux \cite{sase,quevedo,Agha03} or warped \cite{gigupo,agetal} compactifications) will lead to more appealing scenarios.   

{It would be interesting to explore the relation of the present work to string theory compactifications \cite{wi}. For example, compactification of heterotic string theory on
the $T_6/Z_4$ orbifold \cite{grhiko} should, in an appropriate limit, yield a similar set-up to ours (but of course with  a number of additional multiplets).  
However, to our knowledge, a 4d effective description of 6d supergravity subsector of such orbifold string theories has not been given in the literature.}

\vspace{5mm}
\noindent{\large\bf Acknowledgments}
\vspace{5mm}

We thank Wilfried Buchm\"uller for useful comments on the manuscript. 
A.F. thanks high energy theory  groups of KAIST in Daejeon and SNU in Seoul (and especially Kiwoon Choi and Hyung Do Kim) for their hospitality during completion of this work.  

A.F.  was partially supported by the Polish KBN grant 2 P03B 129 24 for years 2003-2005
and by the EC Contract MRTN-CT-2004-503369 - network "The Quest for Unification: Theory
Confronts Experiment" (2004-2008). The  stay of A.F. at DESY is possible owing to Research
Fellowship  granted by  Alexander von Humboldt Foundation.

\appendix
\renewcommand{\theequation}{\Alph{section}.\arabic{equation}}
\setcounter{section}{0}
\setcounter{equation}{0}

\section{Notation and conventions}
\label{a.nc6}

We use the mostly minus metric signature $(+,-,-,-,-,-)$. 
The index conventions are the following: 
6d Einstein indices $A,B,C, \ldots = 0\ldots 3,5,6$,  
6d Lorentz indices $a,b,c, \ldots = 0\ldots 3,5,6$, 
4d Einstein indices $\mu,\nu,\rho, \ldots = 0\ldots 3$,  
4d Lorentz indices $m,n,\ldots = 0\ldots 3$. Furthermore $\alpha,\beta, \ldots = 5,6$. When necessary, we also distinguish Einstein indices with an overdot, e.g. in  $B_{\dot 5 \dot 6}$.  

6d gamma matrices denoted by $\Gamma^{a}$ satisfy 
$\{ \Gamma^{a}, \Gamma^{b} \} =  2 \eta^{ab}$. They are $8 \times 8$ matrices and the 6d spinors  are 8-dimensional. The 4d $4 \times 4$ gamma matrices denoted by $\gamma^{m}$ satisfy  $\{ \gamma^{m}, \gamma^{n} \} = 2 \eta^{m n}$. The convention for $\gamma^5$ is  $\gamma^5 = {\rm diag} (-1,-1,1,1) $ and the chirality projection operators are $P_L = (1 - \gamma^5)/2$, $P_R = (1 + \gamma^5)/2$.  4d spinors are always written in the four-component Dirac notation.   The 4d charge conjugation matrix $C = i \gamma^0 \gamma^2 \gamma^5$ satisfies  
$C^{-1} = C^T = C^\dagger = -C$,  $ C \gamma^{m} C^{-1} = (\gamma^m)^T$.
 
 In our choice of basis the connection between the 6d and 4d  gamma matrices is given by 
\beq
\Gamma^m = \left [ \ba{cc} \gamma^m & 0 \\ 0 & \gamma^m \ea \right ] \, ,
\quad 
\Gamma^5 = \left [ \ba{cc} 0 & i\gamma^5 \\ i\gamma^5 & 0 \ea \right ] \, ,
\quad 
\Gamma^6 = \left [ \ba{cc} 0 & \gamma^5 \\  -\gamma^5 & 0 \ea \right ] \, .
\eeq
Furthermore $\Gamma^A = \Gamma^a  e_a{}^A$,  $\gamma^\mu = \gamma^m  e_m{}^{\mu}$.
 It is also convenient to define 'Lorentz index gravitinos', $\psi_a =   e_a{}^A \psi_A$. 

The fermions in the 6d supergravity action are chiral. The 6d chirality projectors are 
$P= (1 \pm \Gamma^7)/2$ where $\Gamma^7 = {\rm diag} (-1,-1,1,1,1,1,-1,-1)$. The 
gravitino $\psi_A$ and the dilatino $\chi$ have opposite 6d chirality. 
Written in terms of 4d spinors  the 6d fermions  look as  follows
\beq
\psi_\mu =   \left [ \ba{c} \psi_\mu^- \\  \psi_\mu^+ \ea \right ] \, ,
\quad
 \psi_\alpha =   \left [ \ba{c} \psi_\alpha^+ \\  \psi_\alpha^-\ea \right ] \, ,
\quad 
\chi =   \left [ \ba{c} \chi^- \\  \chi^+ \ea\right ] \, .
\eeq
Here  $\psi_\mu^+$, $\psi_\alpha^- $, $\chi^-$ are right-handed chiral 
(in the 4d sense), $P_R \psi = \psi$, while $\psi_\mu^-$, $\psi_\alpha^+ $, $\chi^+$  are left-handed chiral   $P_L \psi = \psi$. 
In the paper we use the following linear combinations of the extra dimensional components of the (Lorentz index) gravitinos:
\beq
\psi_{5+6} \equiv - (\psi_5^+  + i \psi_6^+) \, , 
 \qquad
\psi_{5-6} \equiv - (\psi_5^+ - i \psi_6^+) \, .
\eeq



\begin{thebibliography}{99}
\bibitem{gm}
  A.~H.~Chamseddine, R.~Arnowitt and P.~Nath,
  Phys.\ Rev.\ Lett.\  {\bf 49} (1982) 970.
  R.~Barbieri, S.~Ferrara and C.~A.~Savoy,
  Phys.\ Lett.\ B {\bf 119} (1982) 343.
  L.~J.~Hall, J.~Lykken and S.~Weinberg,
  Phys.\ Rev.\ D {\bf 27} (1983) 2359.
  H.~P.~Nilles,
  Phys.\ Rept.\  {\bf 110}, 1 (1984).
N.~Ohta,
Prog.\ Theor.\ Phys.\  {\bf 70} (1983) 542.

\bibitem{rasu0}
L.~Randall and R.~Sundrum,
Nucl.\ Phys.\ B {\bf 557} (1999) 79
[arXiv:hep-th/9810155].

\bibitem{rovi}
  G.~G.~Ross and O.~Vives,
  Phys.\ Rev.\ D {\bf 67} (2003) 095013
  [arXiv:hep-ph/0211279].
  
\bibitem{lusu}
  M.~A.~Luty and R.~Sundrum,
  Phys.\ Rev.\ D {\bf 62}, 035008 (2000)
  [arXiv:hep-th/9910202].

\bibitem{ghri}
T.~Gherghetta and A.~Riotto,
Nucl.\ Phys.\ B {\bf 623} (2002) 97
[arXiv:hep-th/0110022].

\bibitem{bugago}
I.~L.~Buchbinder, S.~J.~J.~Gates, H.~S.~J.~Goh, W.~D.~.~Linch, M.~A.~Luty, S.~P.~Ng and J.~Phillips,
Phys.\ Rev.\ D {\bf 70} (2004) 025008
[arXiv:hep-th/0305169].

\bibitem{rascst}
R.~Rattazzi, C.~A.~Scrucca and A.~Strumia,
Nucl.\ Phys.\ B {\bf 674} (2003) 171
[arXiv:hep-th/0305184].

\bibitem{grrasc}
T.~Gregoire, R.~Rattazzi, C.~A.~Scrucca, A.~Strumia and E.~Trincherini,
arXiv:hep-th/0411216.
C.~A.~Scrucca,
arXiv:hep-th/0412237.


\bibitem{aa}
A.~Falkowski,
JHEP {\bf 0505} (2005) 073
[arXiv:hep-th/0502072].


\bibitem{noether}
  S.~Ferrara, D.~Z.~Freedman, P.~van Nieuwenhuizen, P.~Breitenlohner, F.~Gliozzi and J.~Scherk,
  Phys.\ Rev.\ D {\bf 15} (1977) 1013.

\bibitem{andigr}
A.~Anisimov, M.~Dine, M.~Graesser and S.~Thomas,
Phys.\ Rev.\ D {\bf 65} (2002) 105011
[arXiv:hep-th/0111235];
A.~Anisimov, M.~Dine, M.~Graesser and S.~Thomas,
JHEP {\bf 0203} (2002) 036
[arXiv:hep-th/0201256].

\bibitem{masc}
  N.~Marcus and J.~H.~Schwarz,
  Phys.\ Lett.\ B {\bf 115} (1982) 111.

\bibitem{nise}
  H.~Nishino and E.~Sezgin,
  Phys.\ Lett.\ B {\bf 144} (1984) 187.

\bibitem{rasase}
  S.~Randjbar-Daemi, A.~Salam, E.~Sezgin and J.~Strathdee,
  Phys.\ Lett.\ B {\bf 151} (1985) 351.
  A.~Salam and E.~Sezgin,
  Phys.\ Scripta {\bf 32} (1985) 283.
  E.~Bergshoeff, T.~W.~Kephart, A.~Salam and E.~Sezgin,
  Mod.\ Phys.\ Lett.\ A {\bf 1} (1986) 267.
  S.~D.~Avramis, A.~Kehagias and S.~Randjbar-Daemi,
  arXiv:hep-th/0504033.

\bibitem{er}
  J.~Erler,
  J.\ Math.\ Phys.\  {\bf 35}, 1819 (1994)
  [arXiv:hep-th/9304104].

\bibitem{howi}
P.~Horava and E.~Witten,
Nucl.\ Phys.\ B {\bf 475} (1996) 94
[arXiv:hep-th/9603142].

\bibitem{mipe}
E.~A.~Mirabelli and M.~E.~Peskin,
Phys.\ Rev.\ D {\bf 58} (1998) 065002
[arXiv:hep-th/9712214].

\bibitem{le}
  H.~M.~Lee,
  arXiv:hep-th/0502093.

\bibitem{lenizu}
H.~M.~Lee, H.~P.~Nilles and M.~Zucker,
Nucl.\ Phys.\ B {\bf 680} (2004) 177
[arXiv:hep-th/0309195].

\bibitem{quevedo}
Y.~Aghababaie, C.~P.~Burgess, S.~L.~Parameswaran and F.~Quevedo,
JHEP {\bf 0303} (2003) 032
[arXiv:hep-th/0212091].

\bibitem{gowi}
  W.~D.~Goldberger and M.~B.~Wise,
  Phys.\ Rev.\ D {\bf 65}, 025011 (2002)
  [arXiv:hep-th/0104170].


\bibitem{popo}
E.~Ponton and E.~Poppitz,
JHEP {\bf 0106} (2001) 019
[arXiv:hep-ph/0105021].

\bibitem{ghgr}
D.~M.~Ghilencea and S.~Groot Nibbelink,
Nucl.\ Phys.\ B {\bf 641} (2002) 35
[arXiv:hep-th/0204094].

\bibitem{racetrack}
N.~V.~Krasnikov,
Phys.\ Lett.\ B {\bf 193}, 37 (1987).
T.~R.~Taylor,
Phys.\ Lett.\ B {\bf 252}, 59 (1990).
J.~A.~Casas, Z.~Lalak, C.~Munoz and G.~G.~Ross,
Nucl.\ Phys.\ B {\bf 347} (1990) 243.
B.~de Carlos, J.~A.~Casas and C.~Munoz,
Nucl.\ Phys.\ B {\bf 399} (1993) 623
[arXiv:hep-th/9204012].



\bibitem{crjusc}
E.~Cremmer, B.~Julia, J.~Scherk, S.~Ferrara, L.~Girardello and P.~van Nieuwenhuizen,
Nucl.\ Phys.\ B {\bf 147} (1979) 105.

\bibitem{gaugino}
  D.~E.~Kaplan, G.~D.~Kribs and M.~Schmaltz,
  Phys.\ Rev.\ D {\bf 62} (2000) 035010
  [arXiv:hep-ph/9911293].
  Z.~Chacko, M.~A.~Luty, A.~E.~Nelson and E.~Ponton,
  JHEP {\bf 0001}, 003 (2000)
  [arXiv:hep-ph/9911323].


\bibitem{kakali}
S.~Kachru, R.~Kallosh, A.~Linde and S.~P.~Trivedi,
Phys.\ Rev.\ D {\bf 68}, 046005 (2003)
[arXiv:hep-th/0301240].

\bibitem{chfani}
K.~Choi, A.~Falkowski, H.~P.~Nilles, M.~Olechowski and S.~Pokorski,
JHEP {\bf 0411}, 076 (2004)
[arXiv:hep-th/0411066].
K.~Choi, A.~Falkowski, H.~P.~Nilles and M.~Olechowski,
arXiv:hep-th/0503216.

\bibitem{chjeok}
  K.~Choi, K.~S.~Jeong and K.~i.~Okumura,
  arXiv:hep-ph/0504037.
  M.~Endo, M.~Yamaguchi and K.~Yoshioka,
  arXiv:hep-ph/0504036.


\bibitem{sase}
A.~Salam and E.~Sezgin,
Phys.\ Lett.\ B {\bf 147} (1984) 47.

\bibitem{Agha03}
  Y.~Aghababaie, C.~P.~Burgess, S.~L.~Parameswaran and F.~Quevedo,
  Nucl.\ Phys.\ B {\bf 680} (2004) 389
  [arXiv:hep-th/0304256].

\bibitem{gigupo}
  G.~W.~Gibbons, R.~Guven and C.~N.~Pope,
  Phys.\ Lett.\ B {\bf 595} (2004) 498
  [arXiv:hep-th/0307238].

\bibitem{agetal}
  Y.~Aghababaie {\it et al.},
  JHEP {\bf 0309} (2003) 037
  [arXiv:hep-th/0308064].

\bibitem{wi}
  E.~Witten,
  Phys.\ Lett.\ B {\bf 155} (1985) 151.

\bibitem{grhiko}
  S.~Groot Nibbelink, M.~Hillenbach, T.~Kobayashi and M.~G.~A.~Walter,
  Phys.\ Rev.\ D {\bf 69} (2004) 046001
  [arXiv:hep-th/0308076].


\end{thebibliography}
\end{document}